\documentclass[AMA,STIX2COL]{MRM}
\articletype{Article Type}%
\usepackage{amssymb}
\usepackage{bbm}
\usepackage{booktabs} % Add this in the preamble if not already there
\usepackage{multirow}
\usepackage{siunitx}
\received{26 April 2016}
\revised{6 June 2016}
\accepted{6 June 2016}
\begin{document}

% due proposte di titoli

%%\title{Voxel-wise supervised IVIM MRI modeling through distributional neural network ensembles}

\title{A Comprehensive Framework for Uncertainty Quantification of Voxel-wise Supervised Models in IVIM MRI}

\author[1,2]{Nicola Casali}{\orcid{0009-0007-6800-5341}}
\author[1]{Alessandro Brusaferri}{\orcid{0000-0003-0838-9826}}
\author[1,2]{Giuseppe Baselli}{\orcid{0000-0003-2978-1704}}
\author[3]{Stefano Fumagalli}{\orcid{0000-0003-3598-6263}}
\author[4]{Edoardo Micotti}{\orcid{0000-0002-5995-2992}}
\author[4]{Gianluigi Forloni}{\orcid{0000-0001-5374-3914}}
\author[4]{Riaz Hussein}{}
\author[1]{Giovanna Rizzo}{\orcid{0000-0002-6341-1304}}
\author[1]{Alfonso Mastropietro}{\orcid{0000-0003-3358-7111}}

\authormark{N. CASALI \textsc{et al}}

\address[1]{\orgname{Istituto di Sistemi e Tecnologie Industriali Intelligenti per il Manifatturiero Avanzato, Consiglio Nazionale delle Ricerche}, \orgaddress{\city{Milan}, \postcode{20133}, \country{Italy}}}
\address[2]{\orgname{Politecnico di Milano}, \orgaddress{\city{Milan}, \postcode{20133}, \country{Italy}}}
\address[3]{\orgdiv{Department of Acute Brain and Cardiovascular Injury}, \orgname{Istituto di Ricerche Farmacologiche Mario Negri IRCCS}, \orgaddress{\city{Milan}, \postcode{20133}, \country{Italy}}}
\address[4]{\orgdiv{Department of Neuroscience}, \orgname{Istituto di Ricerche Farmacologiche Mario Negri IRCCS}, \orgaddress{\city{Milan}, \postcode{20133}, \country{Italy}}}

\corres{Alfonso Mastropietro, Istituto di Sistemi e Tecnologie Industriali Intelligenti per il Manifatturiero Avanzato, Consiglio Nazionale delle Ricerche, Milan, 20133, Italy. \email{alfonso.mastropietro@stiima.cnr.it}}

\abstract[Abstract]{
\fontsize{10pt}{12.5pt}\selectfont
Accurate estimation of intravoxel incoherent motion (IVIM) parameters from diffusion-weighted MRI remains challenging due to the ill-posed nature of the inverse problem and high sensitivity to noise, particularly in the perfusion compartment. In this work, we propose a probabilistic deep learning framework based on Deep Ensembles (DE) of Mixture Density Networks (MDNs), enabling estimation of total predictive uncertainty and decomposition into aleatoric (AU) and epistemic (EU) components. The method was benchmarked against non probabilistic neural networks, a Bayesian fitting approach and a probabilistic network with single Gaussian parametrization. Supervised training was performed on synthetic data, and evaluation was conducted on both simulated and an in vivo dataset.
The reliability of the quantified uncertainties was assessed using calibration curves, output distribution sharpness, and the Continuous Ranked Probability Score (CRPS). MDNs produced more calibrated and sharper predictive distributions for the diffusion coefficient \( D \) and fraction \( f \) parameters, although slight overconfidence was observed in pseudo-diffusion coefficient \( D^* \). The Robust Coefficient of Variation (RCV) indicated smoother in vivo estimates for \( D^* \) with MDNs compared to Gaussian model. Despite the training data covering the expected physiological range, elevated EU in vivo suggests a mismatch with real acquisition conditions, highlighting the importance of incorporating EU, which was allowed by DE.
Overall, we present a comprehensive framework for IVIM fitting with uncertainty quantification, which enables the identification and interpretation of unreliable estimates. The proposed approach can also be adopted for fitting other physical models through appropriate architectural and simulation adjustments.
}


\maketitle

%\backmatter
\section{Introduction}

Intravoxel Incoherent Motion (IVIM) is a diffusion-weighted (DW) MRI technique that enables the simultaneous estimation of tissue diffusion and microvascular perfusion properties \cite{bihan1992capillary,le1988separation,le2019can}. The IVIM model describes the DW signal decay using a bi-exponential function, allowing for the characterization of the microcirculatory perfusion network without the need for contrast agents \cite{wong2024preoperative}. Considering the low end of b-values, IVIM provides quantitative information on three parameters: the diffusion coefficient ($D$), which reflects the slow diffusion of water molecules; the pseudo-diffusion coefficient ($D^*$), associated with the faster, incoherent motion of blood within capillaries; and the perfusion fraction ($f$), representing the proportion of the signal attributed to microvascular perfusion.

Accurate estimation of IVIM parameters is crucial to ensure their reliability in real experimental and clinical scenarios. The IVIM model has shown promising potential in oncology, particularly for evaluating neoangiogenesis and monitoring the efficacy of chemotherapy and radiotherapy. It has been investigated in various anatomical regions, including brain and head tumors, prostate, and breast \cite{le2019can}. Beyond oncology, IVIM has also been applied to skeletal muscle imaging to gain mechanistic insights into disease processes in patient populations \cite{englund2022intravoxel} and to evaluate perfusion during physical exercise \cite{filli2015dynamic,mastropietro2018triggered}.
Moreover, several preclinical studies have demonstrated the utility of mouse models for investigating microvascular alterations and tumor perfusion in the brain using the IVIM model \cite{federau2017intravoxel, iima2014characterization, wu2019evidence}.

Traditional non-linear least squares (LSQ) approaches often exhibit poor performance, particularly in estimating the pseudo-diffusion coefficient ($D^*$), and are highly sensitive to noise~\cite{park2017intravoxel}. Although Bayesian algorithms have demonstrated improved fitting accuracy, they are computationally intensive and rely heavily on empirically defined priors, which may introduce bias in parameter estimation~\cite{while2017comparative}. To address these limitations, Deep Learning (DL) has recently emerged as a promising alternative to conventional methods for estimating IVIM parameters, due to its flexible and powerful mapping capabilities. Several studies have investigated the use of neural networks (NN) within both self-supervised (also known as physics-informed) and fully supervised learning frameworks for IVIM parameter estimation. 

Self-supervised approaches have been explored both voxel-wise — commonly employing Multi-Layer Perceptrons (MLPs) for independent fitting of each voxel~\cite{bertleff2017diffusion, barbieri2020deep, kaandorp2021improved} — and using Convolutional Neural Networks (CNNs), which aim to capture spatial context by learning features across neighboring voxels in the input images~\cite{huang2022unsupervised, vasylechko2022self, huang2023synthetic, luo2025self}. However, these self-supervised methods, particularly CNN-based frameworks, rely critically on access to a substantial and representative amount of in-vivo IVIM data to achieve reliable performance~\cite{wang2025improved}. This dependence may pose a significant practical barrier to widespread application, especially in settings where data collection is costly, limited, or subject to strict constraints.

Supervised learning approaches have shown improved precision compared to self-supervised methods when trained on consistent synthetic datasets~\cite{kaandorp2023deep}. In this context, voxel-wise training strategies using numerical phantoms are commonly exploited ~\cite{mastropietro2022supervised}. 
Attention-based models have been shown to outperform CNNs in estimation accuracy under image-wise settings ~\cite{kaandorp101incorporating}. Nevertheless, a critical limitation of spatially aware supervised models lies in their reliance on training simulations that must properly capture the spatial patterns observed in vivo. This requirement is particularly challenging due to the inherent complexity of tissue heterogeneity and the potential for domain mismatch.

Notably, despite differences in NN architectures, the vast majority of existing studies focus on estimating the point values of IVIM parameters (i.e, approximating the conditional expectation of the latent distribution). A fully distributional regression approach provides a richer and more realistic framework for empirical analysis by enabling a comprehensive characterization of the conditional response distribution beyond single summary statistics such as the mean or median \cite{kneib2023rage}. Importantly, the uncertainty quantification (UQ) afforded by full distributional regression is critical, as it supports the assessment of confidence in parameter estimates \cite{jones2008tractography}, helps characterize the impact of noise, and can guide experimental design by informing optimal sampling strategies and data acquisition protocols \cite{alexander2008general}.

In addition, UQ provides valuable integrated support for assessing IVIM estimation outcomes.
In particular, the parameters $D^*$ and $f$ are often highly sensitive to noise, which can result in unstable fits or non-physiological values. This sensitivity typically arises from the fact that the perfusion compartment contributes minimally to the overall signal decay, especially at higher b-values \cite{le2019can}. Consequently, small fluctuations in the signal, mainly occurring at low b-values, can lead to large variations in parameter estimates.
By quantifying uncertainty, it becomes possible to identify unreliable estimates that could otherwise lead to misinterpretation in clinical or research settings.

To date, despite increasing research interest on UQ in healthcare and medical image analysis~\cite{seoni2023application, huang2024review, lambert2024trustworthy}, the development of DL techniques within a broader distributional regression framework for IVIM remains underexplored and largely overlooked.
To the best of our knowledge, the only exceptions are the works of \cite{zhang2019implicit} and \cite{casali2025physics}: the former employs a MLP-based approach modeling the posterior distribution as a Gaussian with signal-dependent mean and variance, while the latter investigates an ensemble of CNNs within a self-supervised fitting scheme.

Outside of the specific scope of IVIM, UQ has received growing attention in the diffusion MRI (dMRI) community as a valuable tool to enhance model interpretability, robustness, and decision-making. 
For instance, Jallais et al.~\cite{jallais2024introducing} proposed a flexible toolbox leveraging normalizing flows~\cite{papamakarios2021normalizing} to estimate the posterior distribution of different diffusion biophysical models parameters. 
Manzano-Patrón et al.~\cite{manzano2025uncertainty} applied simulation-based inference (SBI) to overcome the computational limitations of traditional Bayesian methods, such as Markov Chain Monte Carlo (MCMC), which often incur significant costs and hinder practical applicability. Their approach enabled the estimation of uncertainty maps for tractography parameters in parametric spherical deconvolution. 
Karimi et al.~\cite{karimi2023deep} introduced a likelihood-free framework based on Mixture Density Networks (MDNs) to recover full posterior distributions in a multi-tensor diffusion model. Similarly, Consagra et al.~\cite{consagra2025deep} employed MDNs for probabilistic regression in the estimation of multi-fiber diffusion parameters, aiming to improve the characterization of brain microstructure.

Notably, the aforementioned studies mainly focus on characterizing the output distribution in the conditioned space, i.e., the aleatoric uncertainty (AU) in the regression task. However, comprehensive uncertainty characterization for IVIM parameter estimation, incorporating both AU and epistemic uncertainty (EU), remains underexplored. EU arises from model limitations or mismatches between training and inference conditions \cite{kendall2017uncertainties} and is essential to evaluate model reliability. In the context of DL regressors, EU corresponds to uncertainty in the NN’s parameters.
For IVIM modeling, distinguishing AU from EU enables clinicians and users to disentangle intrinsic noise-driven variability, particularly affecting sensitive parameters like $D^*$, from cases where real in vivo variability may not be adequately represented in the training data.
Furthermore, quantitative evaluation of uncertainty quality in this domain is limited, with scarce attention given to key metrics such as calibration and sharpness, which assess whether predicted uncertainties are both consistent with observed outcomes and informative.

Leveraging the current state of the art and building upon identified open challenges, the primary objective of this work is to advance the investigation of UQ in DL-based IVIM modeling. To this end, we develop a comprehensive framework for supervised voxel-wise fitting based on a deep ensemble (DE) of distributional NN regressors.
Specifically, we explore both single Gaussian parameterizations and MDNs, the latter providing greater flexibility in modeling complex posterior distributions. Indeed, IVIM parameter distributions may exhibit multimodality, skewness, and heavy tails \cite{zhang2019implicit}, yet an experimental investigation of distributional forms with increasing complexity remains absent in the literature. This work addresses this gap by evaluating more expressive distributional models, with particular emphasis on whether MDNs can improve both prediction accuracy and the reliability of uncertainty estimates. To investigate the reliability of uncertainty estimates, we complement our analysis by adopting standard UQ metrics that quantify two key properties: calibration, which measures the agreement between predicted probabilities and observed frequencies, and sharpness, which captures the concentration of the predicted distributions.

\section{Materials and Methods}
\vspace{1em}
\subsection{Dataset}
\vspace{1em}
\subsubsection{Simulated Data}

To train the proposed approach, 200\,000
 synthetic signals were generated by uniformly sampling IVIM parameters over the following ranges: \(D \in [0, 0.003]~\si{mm^2/s}\),  \(f \in [0, 0.4]\), and \(D^* \in [0.003, 0.2]~\si{mm^2/s}\). The DW signals were synthesized using the same set of \(b\)-values employed in the in vivo acquisitions (see Section~\ref{sec:in_vivo}), according to the IVIM model equation:

\begin{equation}\label{eq:eq_ivim}
S(b) = S_0 \left[ f e^{-bD^*} + (1 - f) e^{-bD} \right],
\end{equation}
where \(S(b)\) denotes the signal at a given DW \(b\) and \(S_0\) is the signal at \(b = 0\). Rician noise was added to the synthetic signals to achieve signal-to-noise ratio (SNR) ranging from 0 to 200, as proposed in Kaandorp et al.\cite{kaandorp2023deep}

For realistic simulation-based testing, $76 \times 76$ pixel
 Shepp-Logan phantoms  were employed following the approach in previous works \cite{mastropietro2022supervised,casali2025physics}, incorporating six distinct regions of interest (ROI) and background noise. Each ROI was assigned a unique set of IVIM parameters, sampled from uniform distributions over the same parameter ranges used during training. A total of 600 phantoms were generated for evaluation, with 200 phantoms each at SNR levels of 25, 50 and 100, to assess the robustness of the network under varying noise conditions. Background pixels were excluded from the analysis.
 
\subsubsection{In Vivo Data}
\label{sec:in_vivo}

To evaluate the approaches trained on simulated data, we employed a preclinical dataset consisting of mouse brain images. This dataset was acquired from mouse brains of animals carrying a genetic mutation in presenilin-1 causing Alzheimer disease or their wild-type littermates. A total of 54 mouse brains were imaged using a 7 Tesla Bruker Biospec 70/30 spectrometer equipped with a transmitter/receiver cryo-coil.  An EPI-DWI sequence was applied along three orthogonal directions using 14 diffusion values:  
b = 0, 15, 60, 100, 150, 170, 190, 220,
280, 440, 560, 700, 850, 1000 s/mm\(^2\). Acquisition parameters were:  
TR = 3 s, TE = 61 ms, 2 segments, 2 repetitions,  
FOV = \(15 \times 15\) mm, MTX = \(76 \times 76\),  
slice thickness = 0.5 mm,  
\(\delta = 5.8\) ms, \(\Delta = 50\) ms. The number of slices on the axial plane was 8 for all acquisitions, resulting in a total of 432 samples across the dataset.
These images were used to evaluate the model, which was trained on simulations, on real in vivo images. 
 
\subsection{Voxel-wise supervised IVIM modeling}

Let \( \mathbf{x} \in \mathbb{R}^b \) denote the input feature vector for a single voxel, where \( b \) is the number of \( b \)-values. Each \( \mathbf{x} \) represents the diffusion signal normalized by the signal at \( b = 0 \). The task is to predict the corresponding IVIM parameters \( \mathbf{y} = [D, D^*, f]^\top \in \mathbb{R}^3 \).

In the supervised setting, we are given a dataset of \( N \) voxel-wise training examples \( \{(\mathbf{x}_i, \mathbf{y}_i)\}_{i=1}^N \), where \( \mathbf{x}_i \) is the normalized input signal and \( \mathbf{y}_i \) the corresponding ground truth IVIM parameters.

The objective is to learn a function \( f_\theta: \mathbb{R}^b \rightarrow \mathbb{R}^3 \), parameterized by \( \theta \), that maps each input \( \mathbf{x}_i \) to a point estimate \( \hat{\mathbf{y}}_i = f_\theta(\mathbf{x}_i) \) of the target parameters.

\subsubsection{MLP-based parameters regression}

A standard Multi-Layer Perceptron (MLP) with two hidden layers used for point regression is defined as:

\begin{equation}
\hat{\mathbf{y}} = f_\theta(\mathbf{x}) = \sigma\left(W_3 \cdot \phi\left(W_2 \cdot \phi\left(W_1 \cdot \mathbf{x} + \mathbf{b}_1\right) + \mathbf{b}_2\right) + \mathbf{b}_3\right),
\end{equation}

where:
\begin{itemize}
    \item \( \phi(\cdot) \) is the elementwise activation function (e.g., exponential linear unit, ELU),
    \item \( \sigma(\cdot) \) is the elementwise sigmoid function to constrain outputs to \([0,1]\),
    \item \( W_1, W_2, W_3 \) are weight matrices, and \( \mathbf{b}_1, \mathbf{b}_2, \mathbf{b}_3 \) are bias vectors,
    \item \( \theta = \{W_i, \mathbf{b}_i\}_{i=1}^3 \) denotes all learnable network parameters.
\end{itemize}
Training this MLP in a supervised regression setting corresponds to solving an optimization problem that minimizes the negative log-likelihood of the data under an assumed noise model. Specifically, if we assume that the target outputs \( \mathbf{y} \) are drawn from a Gaussian distribution with constant, isotropic variance centered at the model output, i.e.,

\begin{equation}
p(\mathbf{y} \mid \mathbf{x}) = \mathcal{N}(\mathbf{y} \mid f_\theta(\mathbf{x}), \sigma^2 \mathbf{I}),
\end{equation}
then the corresponding negative log-likelihood loss becomes:

\begin{equation}
\mathcal{L}_{\text{NLL}}(\theta) = -\log p(\mathbf{y} \mid \mathbf{x}) = \frac{1}{2\sigma^2} \left\| \mathbf{y} - f_\theta(\mathbf{x}) \right\|^2 + \text{const}.
\end{equation}
Minimizing this negative log-likelihood is equivalent (up to a constant) to minimizing the mean squared error (MSE) loss:

\begin{equation}
\mathcal{L}_{\text{MSE}}(\theta) = \frac{1}{N} \sum_{i=1}^N \left\| f_\theta(\mathbf{x}_i) - \mathbf{y}_i \right\|^2.
\end{equation}
This formulation corresponds to maximum likelihood estimation (MLE) under Gaussian noise and results in the model learning to approximate the conditional mean of the target distribution, i.e., \( f_\theta(\mathbf{x}) \approx \mathbb{E}[\mathbf{y} \mid \mathbf{x}] \). However, this point estimation framework does not provide any measure of predictive uncertainty, nor can it capture skewness, multimodal or heteroscedastic characteristics of the conditional distribution \( p(\mathbf{y} \mid \mathbf{x}) \), which are often present in inverse problems like IVIM parameter estimation. These limitations motivate the use of probabilistic models that aim to represent the full predictive distribution rather than only its mean.

\subsubsection{Distributional regression framework}

A straightforward probabilistic extension of the point-estimate model involves representing the conditional distribution over IVIM parameters as a single multivariate Gaussian with input-dependent (heteroscedastic) variance. Specifically, the model assumes:

\begin{equation}
p(\mathbf{y} \mid \mathbf{x}) = \mathcal{N}\left(\mathbf{y} \mid \boldsymbol{\mu}(\mathbf{x}),\boldsymbol{\sigma}^2(\mathbf{x})\right).
\end{equation}
In this setting, following a factorized distribution form, NNs are designed to predict a vector parameterizing the distribution locations \(\boldsymbol{\mu}(\mathbf{x}) \in \mathbb{R}^3\) and a vector of variances \(\boldsymbol{\sigma}^2(\mathbf{x}) \in \mathbb{R}_+^3\), with one mean and one variance related to each IVIM parameter. The model is trained by minimizing the negative log-likelihood of the observed targets under the predicted Gaussian distribution.

To model more complex distributions \( p(\mathbf{y} \mid \mathbf{x}) \), this single Gaussian can be generalized to a mixture of Gaussians. This leads to the MDN \cite{bishop1994mixture}, where the conditional distribution is modeled as:

\begin{equation}
p(\mathbf{y} \mid \mathbf{x}) = \sum_{k=1}^{K} \pi_k(\mathbf{x}) \, \mathcal{N}\left(\mathbf{y} \mid \boldsymbol{\mu}_k(\mathbf{x}), \boldsymbol{\sigma}^2_k(\mathbf{x})\right).
\label{eq:mdn}
\end{equation}
Here, each component \(k = 1, \dots, K\) is parameterized by a mixture weight \(\pi_k(\mathbf{x}) \in (0, 1)\) (with \(\sum_k \pi_k(\mathbf{x}) = 1\)), a mean vector \(\boldsymbol{\mu}_k(\mathbf{x}) \in \mathbb{R}^3\), and a variance vector \(\boldsymbol{\sigma}^2_k(\mathbf{x}) \in \mathbb{R}_+^3\). All parameters are predicted by a NN conditioned on the input \(\mathbf{x}\) using the negative log-likelihood of Eq.\ref{eq:mdn} as loss function.

\begin{figure*}[!t]
\centerline{\includegraphics[width=0.85\textwidth]{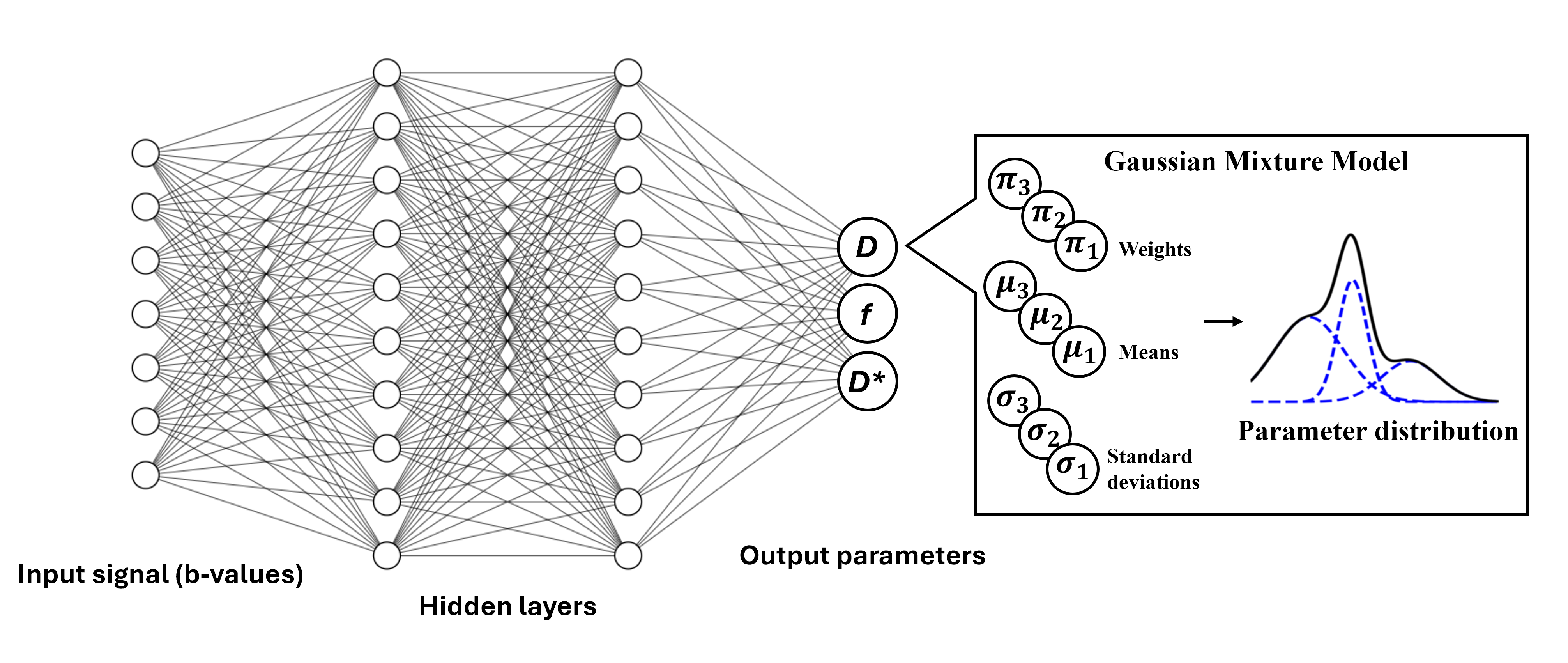}}
\caption{Mixture Density Network architecture. A fully connected NN takes the IVIM signal as input, with one input neuron per b-value. For each IVIM parameter, the network outputs the parameters of a Gaussian Mixture Model, specifically the weights, means, and standard deviations, corresponding to the chosen number of mixture components. In this case, it is shown an example with three components.}
\label{fig:net}
\end{figure*}

\subsubsection{Characterizing epistemic uncertainty via Deep Ensembles}
\label{sec:ensemble}
The heteroscedastic models discussed above capture AU-that is, the inherent noise or variability in the data, which cannot be reduced even with additional observations. However, these models are agnostic to EU \cite{gustafsson2020evaluating}, which arises from limited data and reflects the model’s uncertainty about its own parameters. EU represents the model’s lack of knowledge, which may stem from suboptimal model parameterizations or from encountering inputs that lie outside the distribution seen during training \cite{lohr2024towards}.

A principled way to incorporate EU in NNs is through Bayesian inference \cite{neal1996bayesian}. In this framework, the network weights \( \theta \) are treated as random variables governed by a posterior distribution \( p(\theta \mid \mathcal{D}) \), which captures the uncertainty over model parameters given the observed dataset \( \mathcal{D} \). According to Bayes’ theorem, this posterior is defined as:

\begin{equation}
p(\theta \mid \mathcal{D}) = \frac{p(\mathcal{D} \mid \theta) \, p(\theta)}{\int_{\theta} p(\mathcal{D} \mid \theta) \, p(\theta) \, d\theta}.
\end{equation}
Here, \( p(\mathcal{D} \mid \theta) \) is the likelihood of the data given the parameters, \( p(\theta) \) is the prior over the network weights, and the denominator represents the evidence, which normalizes the posterior.

During testing, the predictive distribution for a new input \(\mathbf{x}_*\) is obtained by marginalizing over the posterior distribution of the network weights:
\begin{equation}
p(\mathbf{y}_* \mid \mathbf{x}_*, \mathcal{D}) = \int_{\Omega} p(\mathbf{y}_* \mid \mathbf{x}_*, \boldsymbol{\theta}) \, p(\boldsymbol{\theta} \mid \mathcal{D}) \, d\boldsymbol{\theta}.
\end{equation}
This expression accounts for the uncertainty in the model parameters by integrating over all plausible configurations of the network weights. However, exact computation of this integral is generally intractable for deep NNs. Therefore, approximate methods—such as variational inference, Monte Carlo Dropout, or DE—have been proposed in the Deep Learning literature to achieve computationally tractable posterior approximations \cite{jospin2022hands}.

To incorporate EU, we chose to adopt a DE approach \cite{lakshminarayanan2017simple}, which provides a practical and effective way to approximate the Bayesian posterior over NN weights without requiring complex variational approximations or modifications of the architecture.
In this setting, \( M \) NNs \( \{f_{\theta^{(m)}}\}_{m=1}^M \) are independently trained with different random initializations. Formally for the MDN (and for the single Gaussian model by imposing \(K=1\)), the conditional distribution for each member of the ensemble is:

\begin{equation}
p_m(\mathbf{y} \mid \mathbf{x}) = \sum_{k=1}^K \pi_k^{(m)}(\mathbf{x}) \, \mathcal{N}\left(\mathbf{y} \mid \boldsymbol{\mu}_k^{(m)}(\mathbf{x}), \boldsymbol{\sigma}_k^{2\,(m)}(\mathbf{x})\right)
\label{eq:mdn_ensemble}
\end{equation}
The final predictive distribution is obtained as a mixture with uniform weights over the ensemble members:
\begin{equation}
p(\mathbf{y} \mid \mathbf{x}) = \frac{1}{M} \sum_{m=1}^M p_m(\mathbf{y} \mid \mathbf{x}).
\end{equation}

To evaluate the uncertainty metrics described in Section~\ref{sec:uq_metrics}, we constructed an empirical predictive distribution for each test input using an ensemble of \( M = 5 \) trained probabilistic networks. For each test point \(\mathbf{x}\), we sampled \( S = 100 \) outputs from the predicted mixture distribution \( p_m(\mathbf{y} \mid \mathbf{x}) \) of each ensemble member. Pooling all samples across ensemble members yielded an empirical predictive distribution with \( M \times S = 500 \) samples per test point.

\subsubsection{Disentangling aleatoric and epistemic uncertainties}

The use of DE in conjunction with probabilistic models enables the decomposition of predictive uncertainty into AU and EU. This separation can be formalized using the law of total variance \cite{depeweg2018decomposition}, which states:

\begin{equation}
\operatorname{Var}(\mathbf{y} \mid \mathbf{x}) = \underbrace{\mathbb{E}_{m}[\operatorname{Var}(\mathbf{y} \mid \mathbf{x}, f_{\theta^{(m)}})]}_{\text{Aleatoric Uncertainty}} + \underbrace{\operatorname{Var}_{m}(\mathbb{E}[\mathbf{y} \mid \mathbf{x}, f_{\theta^{(m)}}])}_{\text{Epistemic Uncertainty}}
\label{eq:total_variance}
\end{equation}

Here, \( f_{\theta^{(m)}} \) denotes the \( m \)-th model in the ensemble. Intuitively, the AU is assessed through the expected predictive variance across the ensemble (i.e., the average of the individual model variances). EU, on the other hand, is quantified as the variance in the predicted means across ensemble members.

Considering the MDN conditional distribution of Eq.~\eqref{eq:mdn_ensemble}, we can approximate the predictive mean and variance of each model \( m \) using the mixture moments:

\begin{align}
\mathbb{E}[\mathbf{y} \mid \mathbf{x}, f_{\theta^{(m)}}] 
&= \sum_{k=1}^K \pi_k^{(m)}(\mathbf{x}) \boldsymbol{\mu}_k^{(m)}(\mathbf{x}), \\[6pt]
\operatorname{Var}(\mathbf{y} \mid \mathbf{x}, f_{\theta^{(m)}}) 
&= \sum_{k=1}^K \pi_k^{(m)}(\mathbf{x}) \Big[
    \boldsymbol{\sigma}_k^{2\,(m)}(\mathbf{x}) \notag \\
&\quad + \left(
        \boldsymbol{\mu}_k^{(m)}(\mathbf{x}) 
        - \mathbb{E}[\mathbf{y} \mid \mathbf{x}, f_{\theta^{(m)}}]
    \right)^2
\Big].
\end{align}

Once the per-model moments are computed, we apply Eq.~\eqref{eq:total_variance}, aggregating across ensemble members:

\begin{align}
\text{AU} &= \frac{1}{M} \sum_{m=1}^M \operatorname{Var}(\mathbf{y} \mid \mathbf{x}, f_{\theta^{(m)}}), \\
\text{EU} &= \frac{1}{M} \sum_{m=1}^M \left( \mathbb{E}[\mathbf{y} \mid \mathbf{x}, f_{\theta^{(m)}}] - \bar{\boldsymbol{\mu}}
 \right)^2,
\end{align}
where \( \bar{\boldsymbol{\mu}}
\) is the ensemble mean:
\begin{equation}
\bar{\boldsymbol{\mu}}
 = \frac{1}{M} \sum_{m=1}^M \mathbb{E}[\mathbf{y} \mid \mathbf{x}, f_{\theta^{(m)}}].
\end{equation}

At inference time, both AU and EU were reported as standard deviations by taking the square root of the corresponding variance estimates \cite{jallais2024introducing}, ensuring that the uncertainty values retained the same physical units as the predicted IVIM parameters. To enable fair comparison across parameters with different magnitudes and units, uncertainty estimates were normalized by the prior range of each IVIM parameter used in the synthetic training dataset.

\subsection{Experimental procedure}

The probabilistic MDN (Figure \ref{fig:net}) and a probabilistic network with a single Gaussian component (hereafter referred to as Gaussian) were implemented and compared to a standard MLP based regression. The predictions from these NNs were also compared with those obtained using a state-of-the-art Bayesian fitting method \cite{gustafsson2018impact}. MDNs with different numbers of mixture components (2, 3, 5, 10, and 20) were assessed, treating the number of components as a hyperparameter. The optimal configuration was selected based on the best trade-off between validation performance and model complexity, as evaluated by the mean validation loss averaged over five independent training runs.

All models shared the same architecture consisting of two hidden layers with ELU activation functions. The number of hidden neurons in each hidden layer was set equal to 64. The network architecture was selected based on previous studies that optimized hyperparameters for the IVIM fitting task \cite{kaandorp2023deep,kaandorp2021improved}. 

The input to the models was the IVIM signal sampled at each \(b\)-value, normalized by the signal at \(b = 0\). The target parameters were normalized to the interval \([0, 1]\) using the minimum and maximum values of the simulation ranges. A sigmoid activation function was applied to the output layer to ensure that the network predictions remained within this normalized range. During inference, the predicted values were rescaled to the original physiological ranges of the IVIM parameters.

The simulated dataset was divided into 80\% training and 20\% validation subsets.
Each model configuration was trained five times with different random weight initializations, as explained in Section \ref{sec:ensemble}. The training used Adam optimizer \cite{adam2014method} with a learning rate of 
\(10^{-4}\), a batch size of 128, and ran for 1000 epochs. The MLP was trained using MSE as the loss function, while the MDN and Gaussian were optimized using the negative log-likelihood, computed as the negative logarithm of Eq.~\eqref{eq:mdn}.

All experiments were implemented in the PyTorch framework and executed on a NVIDIA GeForce RTX 3070 Ti GPU. 

 Since both the MDN and Gaussian models output full probability distributions, point estimates were obtained using the Maximum a Posteriori (MAP) principle, specifically, by selecting the mean of the Gaussian component with the highest mixture weight. This choice was made following other works about probabilistic regression in dMRI \cite{jallais2024introducing,consagra2025deep}.

\subsection{Model evaluation}
\vspace{1em}
\subsubsection{Accuracy metrics}

To evaluate the accuracy of parameter estimates on the simulated test set of phantoms, the relative median absolute error (MdAE) and the relative median bias (MdB) were calculated as follows: 

\begin{equation}
\text{MdAE} = \text{median} \left( \frac{|\mathbf{\theta}_{\text{pred}} - \mathbf{\theta}_{\text{true}}|}{\mathbf{\theta}_{\text{true}}} \right).
\end{equation}

\begin{equation}
\text{MdB} = \text{median} \left( \frac{\mathbf{\theta}_{\text{pred}} - \mathbf{\theta}_{\text{true}}}{\mathbf{\theta}_{\text{true}}} \right).
\end{equation}

To assess precision and repeatability within homogeneous ROI, a robust analog of the coefficient of variation (RCV) was employed \cite{arachchige2022robust}.

\begin{equation}
\text{RCV} = \text{1.486} \left( \frac{\text{MAD}_{\theta_\text{pred}}}{\text{median}_{\theta_\text{pred}}} \right).
\end{equation}

The RCV was calculated separately for each region of the phantom, where MAD refers to the median absolute deviation of the estimated parameters. The median RCV across all simulations was then reported.

\subsection{Calibration and sharpness evaluation metrics}
\label{sec:uq_metrics}

To evaluate the quality and reliability of the predicted uncertainty estimates, we employed well-established calibration and sharpness metrics alongside proper scoring rules. Since these uncertainty metrics require ground truth values for meaningful assessment, the evaluation was performed on the simulated test set.

Calibration metrics measure the agreement between predicted uncertainties and the observed error distribution by assessing whether predicted confidence intervals (CIs) contain the true values at the expected frequencies~\cite{kuleshov2018accurate,yao2019quality,sluijterman2024evaluate,laves2020well}. Sharpness, in contrast, quantifies the concentration or narrowness of the predictive distributions and depends solely on the forecasts themselves~\cite{gneiting2007probabilistic}.

Proper scoring rules, such as the Continuous Ranked Probability Score (CRPS), provide a unified quantitative framework that captures both calibration and sharpness \cite{gneiting2007strictly}. While calibration reflects the reliability of uncertainty estimates—indicating how well predicted uncertainties correspond to observed errors—sharpness rewards predictions with tighter, more precise distributions.

From this empirical distribution, we computed predictive quantiles at various nominal levels (e.g., 5\%, 10\%, \ldots, 95\%) to construct Prediction Intervals (PIs). For a nominal level \( \gamma \)\%, the PI for input \( x_i \) is defined as:

\begin{equation}
\mathrm{PI}_\gamma(x_i) =
\left[
Q_{x_i}\left(\frac{100 - \gamma}{2}\%\right),
\;
Q_{x_i}\left(\frac{100 + \gamma}{2}\%\right)
\right],
\end{equation}

where \( Q_{x_i}(p) \) is the empirical \( p \)-percentile of the predictive samples for \( x_i \). 

We then assessed Prediction Interval Coverage Probability (PICP), which measures the proportion of true values falling within the corresponding PIs~\cite{yao2019quality,kuleshov2018accurate,sluijterman2024evaluate}:

\begin{equation}
\mathrm{PICP}(\gamma) = \frac{1}{N_{\text{test}}} \sum_{i=1}^{N_{\text{test}}} \mathbbm{1}\left\{ y_i \in \mathrm{PI}_\gamma(x_i) \right\},
\end{equation}

where:
\begin{itemize}
  \item \( N_{\text{test}} \) is the total number of test samples,
  \item \( y_i \) is the true target value for the \( i \)-th sample,
  \item \( \mathbbm{1}\{\cdot\} \) is the indicator function, which is 1 if the condition is true and 0 otherwise.
\end{itemize}

Ideally, \( \mathrm{PICP}(\gamma) \) should closely match the nominal confidence level \( \gamma \); for instance, a 90\% prediction interval should contain the true target value approximately 90\% of the time. To assess the calibration properties of each probabilistic model, we computed \( \mathrm{PICP} \) at various nominal levels and constructed calibration plots. We then quantified the miscalibration area, defined as the absolute area between the ideal calibration line and the empirical PICP curve. A smaller miscalibration area indicates better calibration, with a value approaching zero reflecting near-perfect reliability of the predicted uncertainty intervals.

To evaluate the sharpness of the prediction intervals, we also used the Prediction Interval Normalized Average Width (PINAW). This metric computes the average width of the PIs, normalized by the range of the target variable, denoted by \( R \). Lower PINAW values indicate narrower (i.e., sharper) intervals:

\begin{equation}
\mathrm{PINAW}(\gamma) = \frac{1}{N_{\text{test}} \cdot R} \sum_{i=1}^{N_{\text{test}}} \left( U_\gamma(x_i) - L_\gamma(x_i) \right),
\end{equation}
where \( U_\gamma(x_i) \) and \( L_\gamma(x_i) \) are the upper and lower bounds of the \( \gamma \)\% PI for input \( x_i \), and \( R \) is the range of the ground-truth target variable across the test set.
We report the conventional PINAW at the 90\% nominal confidence level (PINAW$_{90}$),  corresponding to the central 90\% prediction interval \cite{khosravi2011comprehensive}.

Finally, we computed the Continuous Ranked Probability Score (CRPS), a proper scoring rule that evaluates the entire predictive cumulative distribution function (CDF) rather than just interval estimates. For a predictive CDF \( F \) and observation \( y \), CRPS is defined as:

\begin{equation}
\mathrm{CRPS}(F, y) = \int_{-\infty}^{\infty} \left( F(z) - \mathbbm{1}\{ y \leq z \} \right)^2 dz,
\end{equation}
where \( \mathbbm{1}\{ y \leq z \} \) is the empirical CDF of the observed value. CRPS generalizes the mean absolute error for probabilistic forecasts, rewarding predictions that are both accurate and well-calibrated. Lower CRPS values indicate better overall predictive performance.
\label{sec:uq_metrics}

\subsubsection{In vivo analysis}

For the in vivo dataset, IVIM analysis was restricted to an anatomically defined ROI in the cerebral cortex.  For each subject, a manual ROI was drawn on the \(b=0\) images to include the cortical ribbon while excluding cerebrospinal fluid and large vessels.  Voxel‑wise estimates of \(D\), \(f\), and \(D^*\) were then obtained using each modeling approach (Bayesian, MDN, Gaussian, MLP).  Within this ROI, we computed the median and RCV for each parameter per subject and aggregated these statistics across the test set.

\section{Results}

The number of components for the MDN was selected based on the mean validation loss across five repetitions for varying numbers of components. A 10-component MDN provided the optimal balance between model complexity and predictive performance and was therefore selected for all subsequent experiments.

\subsection{Simulated Data}

\subsubsection{Accuracy}

\begin{figure*}[!t]
\centerline{\includegraphics[width=\textwidth]{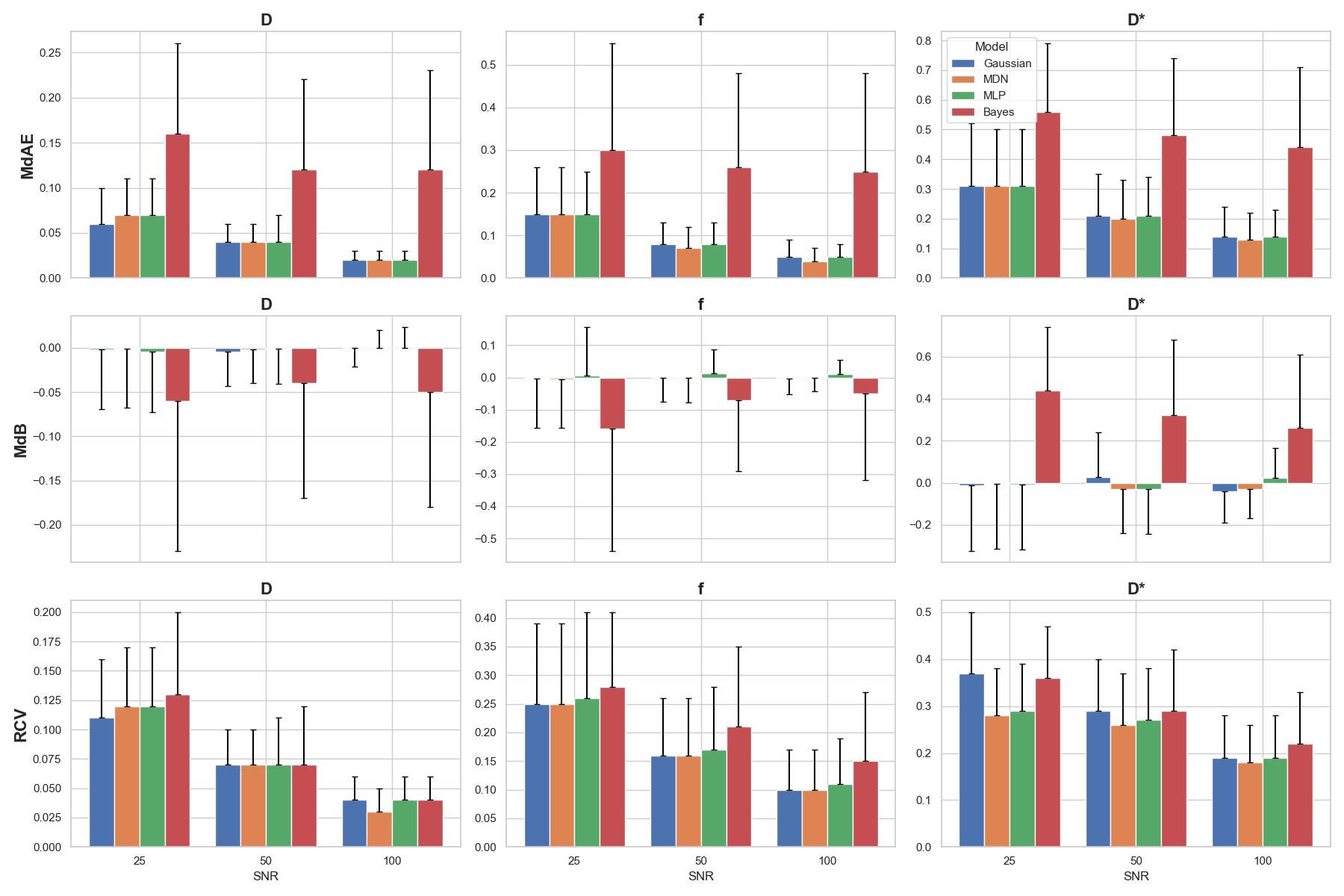}}
\caption{Barplots showing the performance of different tested architectures. The three metrics evaluated on simulated phantoms-MdAE, MdB, and RCV-are displayed for each model, grouped by three SNR levels (25, 50, and 100). Results represent the median values with uncertainty indicated by the median absolute deviation (MAD).}
\label{fig:bar}
\end{figure*}

The values of the metrics MdAE, MdB, and RCV are reported in Figure~\ref{fig:bar}, evaluated on the simulated test set at different SNR levels.  
NNs performed similarly in terms of MdAE, all outperforming the Bayesian model across all three parameters. Bayes consistently showed the highest MdAE across all SNR levels. A similar trend was observed for MdB, where the NNs achieved comparable performance and the Bayesian model again exhibited the worst results.

For RCV, the MDN model achieved the lowest values for \( D^* \) at all noise levels. Specifically, \( 28.5 \pm 10.2 \%\), \( 25.8 \pm 10.6 \%\), and \( 17.7 \pm 8.4\% \) for SNR 25, 50, and 100, respectively. For \( f \) and \( D \), the NNs had similar RCV values, while Bayes performed slightly worse.

\subsubsection{Analysis of the predicted uncertainties}

\begin{figure*}[!t]
\centerline{\includegraphics[width=0.8\textwidth]{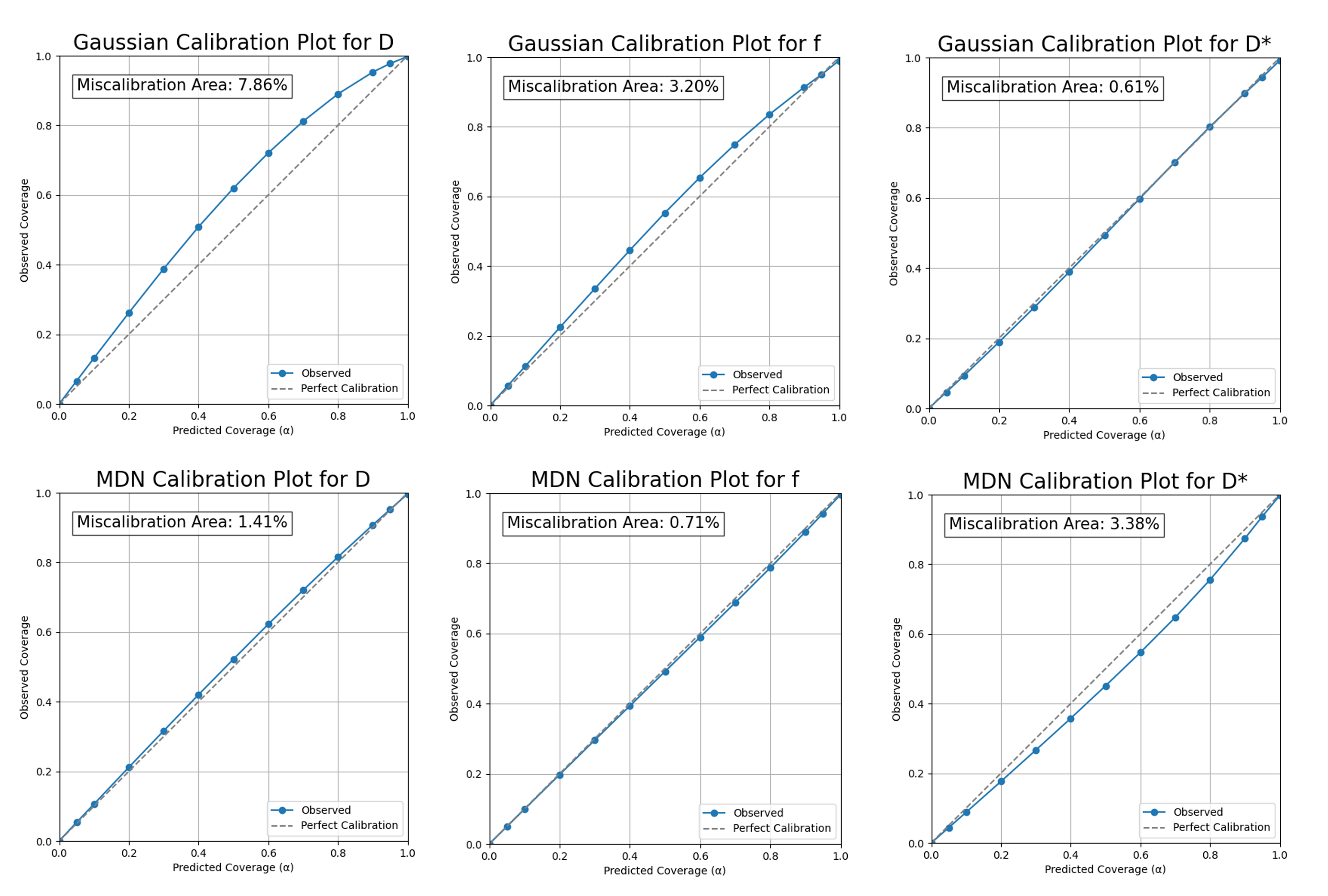}}
\caption{Calibration plots for Gaussian and MDN models for each parameter on the simulated test set. The calibration plots are calculated at PICP at different nominal levels. The miscalibration area measures the deviation of the observed coverage from the ideal one.}
\label{fig:calibration}
\end{figure*}

Figure~\ref{fig:calibration} presents the calibration plots for the Gaussian and MDN models. The MDN demonstrated in general better calibration on the \( D \) and \( f \) parameters, as reflected by the lower miscalibration areas (1.41\% vs. 7.86\% for \( D \), and 0.71\% vs. 3.20\% for \( f \)). In contrast, the Gaussian model achieved superior calibration on \( D^* \), exhibiting a lower miscalibration area compared to MDN (0.61\% vs. 3.38\%). The MDN calibration plot for \( D^* \) demonstrated a slightly overconfidence, possibly due to its greater representational flexibility in modeling complex conditional distributions.

 We report in Table~\ref{tab:uq_metrics} a quantitative comparison of predictive uncertainty using the CRPS and PINAW$_{90\%}$ metrics for each IVIM parameter. Notably, for MDN the CRPS for \( D \) was reduced from 3.4 to 3.0 (\( \times 10^{-5}~\mathrm{mm}^2/\mathrm{s} \)), and for \( f \) from 1.1\% to 1.0\%. For \( D^* \), both models yielded comparable CRPS values.

Additionally, the PINAW$_{90\%}$ analysis revealed that the MDN produced narrower 90\% prediction intervals for both \( D \) and \( f \), which aligns well with the general goal of maximizing sharpness while maintaining calibration \cite{gneiting2007probabilistic}. Specifically, PINAW$_{90}$ was reduced from 10.5\% to 7.6\% for \( D \), and from 21.1\% to 17.2\% for \( f \). For \( D^* \), both models produced wider intervals (around 52\%), reflecting the inherent challenge in estimating this parameter.

\begin{table*}[ht]
\centering
\resizebox{\textwidth}{!}{  % auto-resizes height to maintain aspect ratio
\begin{tabular}{lccc|ccc|ccc}
\hline
 & \multicolumn{3}{c|}{\textbf{D}} & \multicolumn{3}{c|}{\textbf{f}} & \multicolumn{3}{c}{\textbf{D*}} \\
  \textbf{Model}            & \textbf{CRPS} ($10^{-5} ~\mathrm{mm}^2/\mathrm{s}$) & \textbf{PINAW$_{90\%}$}  &     & \textbf{CRPS} (\%) & \textbf{PINAW$_{90\%}$} &     & \textbf{CRPS} ($10^{-2}~\mathrm{mm}^2/\mathrm{s}$) & \textbf{PINAW$_{90\%}$}  &     \\ \hline
Gaussian      & 3.4 (1.7)        & 0.11 (0.04) &     & 1.1 (0.5) & 0.21 (0.07)  &     & 1.3 (0.7)        & 0.52 (0.19) &     \\
MDN           &  \textbf{3.0 (1.7)}       &  \textbf{0.08 (0.03)}  &     &  \textbf{1.0 (0.5)} &  \textbf{0.17 (0.07)}  &     & 1.3 (0.8)        &  \textbf{0.52 (0.21)} &     \\
\hline
\end{tabular}
}
\caption{Comparison of Gaussian and MDN models on CRPS and PINAW$_{90}$ on the simulated test set. Values are median (MAD). }
\label{tab:uq_metrics}
\end{table*}

%\subsubsection{AU and EU components}

\begin{figure*}[!t]
\centerline{\includegraphics[width=0.7\textwidth]{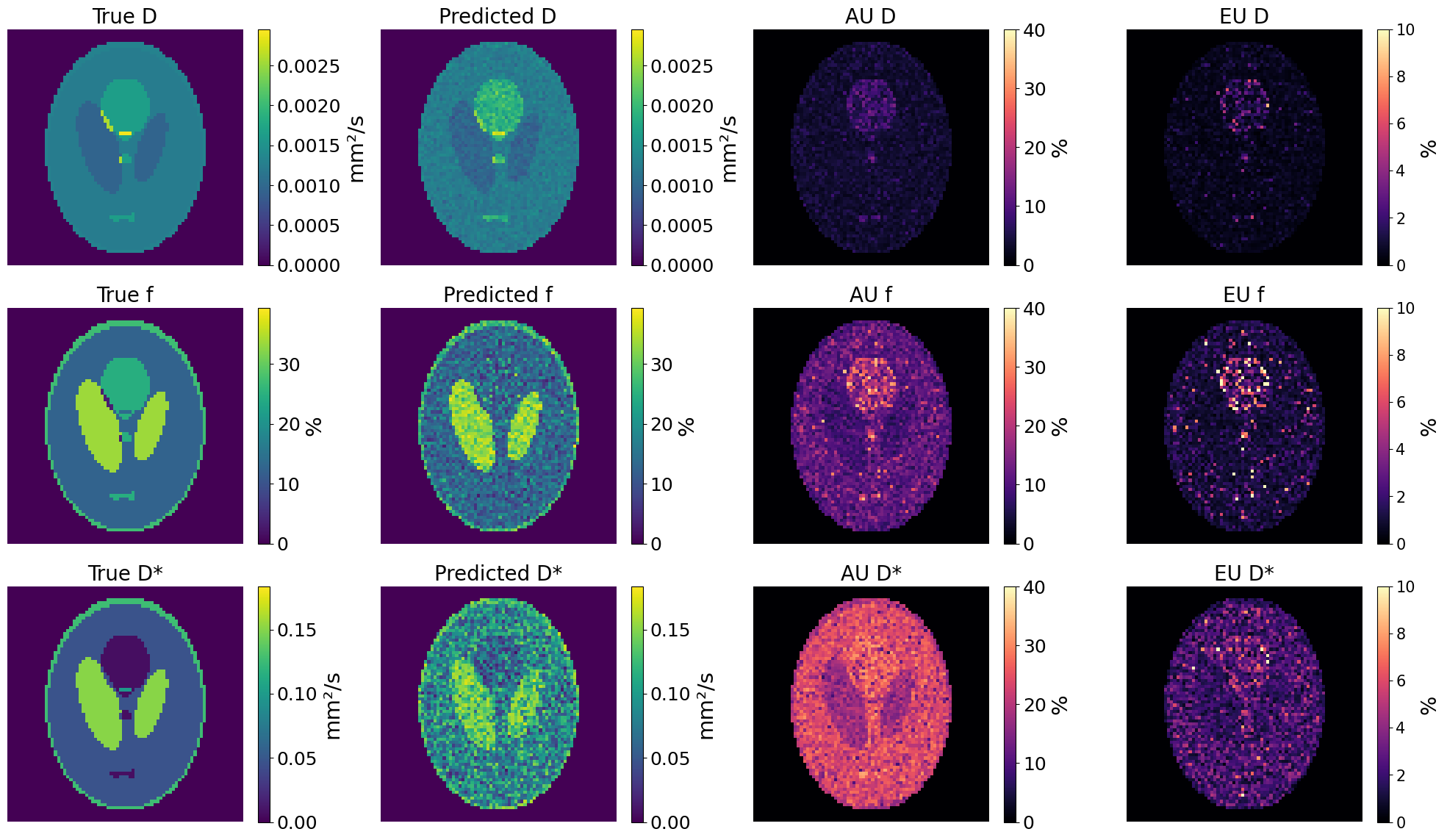}}
\caption{An example of prediction of MDN on a simulated phantom with SNR 25. The image shows the prediction of the network along with the aleatoric (AU) and epistemic (EU) uncertainty maps.}
\label{fig:simulations}
\end{figure*}

Figure~\ref{fig:simulations} shows an example of a simulated phantom at SNR~25, highlighting the disentangled AU and EU maps. The qualitative maps illustrate that \( D^* \) is the parameter with the highest uncertainty, followed by \( f \) and \( D \). The dominant contribution arises from AU, which is attributed to noise in the data, whereas EU is substantially lower in comparison.

These qualitative assessments are complemented by Table~\ref{tab:combined_uq}, which reports the decomposed predictive uncertainties for both the Gaussian and MDN models at different SNR levels on the full simulated test set. As expected, AU decreased as SNR increased for both models, consistent with reduced observational noise. EU remained markedly lower than AU, indicating that the test simulations were well aligned with the distribution of the training data. Among the IVIM parameters, \( D^* \) consistently exhibited the highest uncertainty across models, followed by \( f \), whereas \( D \) was estimated with the greatest confidence by the networks.

\begin{table*}[ht]
\centering
\begin{tabular}{clllllllll}
\hline
\textbf{Type} & \textbf{SNR} & \textbf{Model} & \multicolumn{3}{c}{\textbf{AU (\%)}} & \multicolumn{3}{c}{\textbf{EU (\%)}} \\
              &              &                & \textbf{D} & \textbf{f} & \textbf{D*} & \textbf{D} & \textbf{f} & \textbf{D*} \\
\hline
\multirow{6}{*}{Simulated}
& \multirow{2}{*}{25}  & Gaussian & 4.6 (1.5)  & 9.7 (2.0)  & 20.7 (4.9)  & 1.3 (0.5)  & 1.8 (0.7)  & 3.5 (1.4) \\
&                       & MDN      & 4.5 (1.5)  & 10.5 (2.2) & 21.1 (4.0)  & 0.5 (0.2)  & 1.2 (0.4)  & 2.0 (0.6) \\
& \multirow{2}{*}{50}  & Gaussian & 2.8 (0.9)  & 5.5 (1.0)  & 13.8 (4.6)  & 0.9 (0.3)  & 1.3 (0.5)  & 2.3 (0.9) \\
&                       & MDN      & 2.3 (0.8)  & 5.3 (1.1)  & 14.1 (4.8)  & 0.3 (0.1)  & 0.6 (0.2)  & 1.5 (0.6) \\
& \multirow{2}{*}{100} & Gaussian & 2.4 (0.7)  & 4.2 (0.7)  & 11.3 (4.2)  & 0.9 (0.3)  & 1.2 (0.6)  & 2.6 (1.1) \\
&                       & MDN      & 1.6 (0.5)  & 3.5 (0.4)  & 10.8 (4.9)  & 0.3 (0.1)  & 0.5 (0.2)  & 1.2 (0.5) \\
\hline
\multirow{2}{*}{In vivo}
&                       & Gaussian & 2.1 (0.3)  & 6.3 (1.1)  & 35.6 (4.2)  & 0.8 (0.2)  & 3.0 (0.5)  & 6.6 (1.8) \\
&                       & MDN      & 2.2 (0.9)  & 6.5 (1.8)  & 30.6 (1.0)  & 0.3 (0.1)  & 0.7 (0.1)  & 2.3 (1.0) \\
\hline
\end{tabular}
\caption{Comparison of AU and EU for the Gaussian and MDN models on simulated phantoms (at various SNR levels) and on in vivo brain data. Values are reported as median (MAD).}
\label{tab:combined_uq}
\end{table*}

\subsection{In Vivo Data}
\vspace{1em}
\subsubsection{IVIM Parameters Estimation}

\begin{figure*}[!t]
\centerline{\includegraphics[width=0.75\textwidth]{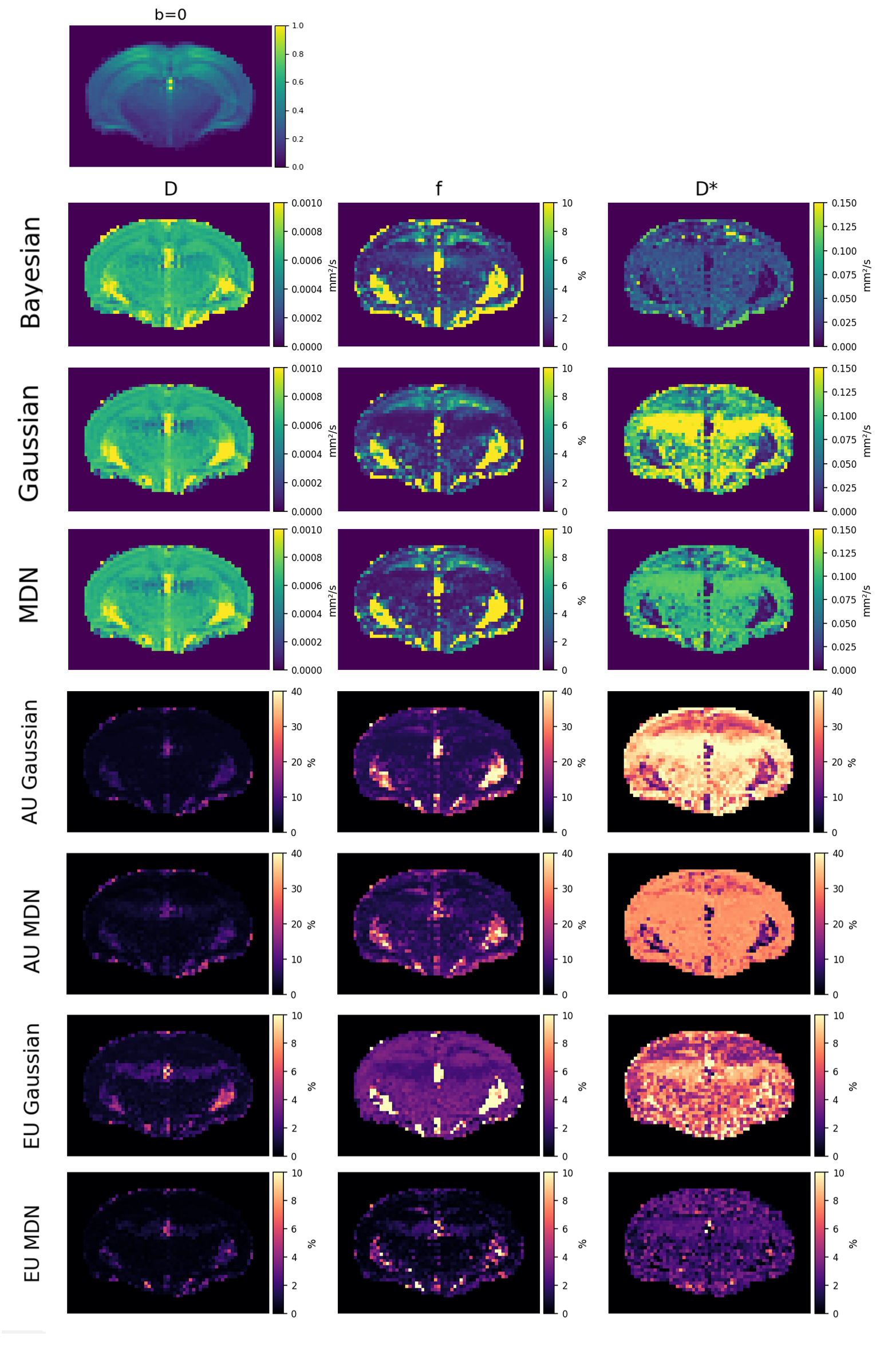}}
\caption{Bayesian, MDN, and Gaussian predictions of the three IVIM parameters on in vivo mouse brain data are presented. The top row shows the b=0 image for the same slice. For the two probabilistic models, Gaussian and MDN, the uncertainty maps are shown separately for aleatoric (AU) and epistemic (EU) components.}
\label{fig:vivo}
\end{figure*}

An example of predictions on the in vivo brain mouse is reported in Figure \ref{fig:vivo}. 
Table~\ref{tab:vivo_val} presents the median and MAD of IVIM parameter estimates across the four different models. All methods produced similar estimates for \(D\) (\(\sim 6.2\text{--}6.3 \times 10^{-4}\) mm\textsuperscript{2}/s). There was slightly variability in \(f\), where the Bayesian model yielded the lowest values (\(1.2 \pm 0.2\%\)), while Gaussian and MLP estimated higher values (\(\sim 2.2\text{--}2.3\%\)). NNs also reported higher pseudo-diffusion values \(D^*\) (\(\sim 0.09\) mm\textsuperscript{2}/s) compared to the Bayesian model (\(\ 0.04\) mm\textsuperscript{2}/s) .

Table~\ref{tab:rcv} reports the RCV in brain ROIs. In this case, all models achieved low RCV for \(D\) (\(\sim4.5\%\)), while variability in \(f\) is the highest for Bayesian (\(55.8\%\)) and lowest for MLP (\(37.9\%\)). MDN attained the best value for \(D^*\) (\(16.0\%\)), compared to Bayesian (\(30.0\%\)).
\begin{table}[htbp]
\resizebox{\columnwidth}{!}{%
\begin{tabular}{lccc}
\toprule
\textbf{Model} & \textbf{D ($\times 10^{-4}$ mm\textsuperscript{2}/s)} & \textbf{f (\%)} & \textbf{D\textsuperscript{*} (mm\textsuperscript{2}/s)} \\
\midrule
Bayesian & $6.2 \pm 0.1$ & $1.2 \pm 0.2$ & $0.04 \pm 0.002$ \\
MDN      & $6.3 \pm 0.1$ & $1.4 \pm 0.3$ & $0.09 \pm 0.003$ \\
Gaussian & $6.2 \pm 0.1$ & $2.2 \pm 0.3$ & $0.09 \pm 0.01$ \\
MLP      & $6.3 \pm 0.1$ & $2.3 \pm 0.3$ & $0.08 \pm 0.01$ \\
\bottomrule
\end{tabular}%
}
\caption{Median (MAD) of values for $D$, $f$, and $D^*$ estimated in a ROI within the mouse brain test set, using Bayesian inference, MDN, Gaussian model, and MLP.}
\label{tab:vivo_val}
\end{table}

\begin{table}[htbp]
\resizebox{\columnwidth}{!}{%
\begin{tabular}{lccc}
\toprule
\textbf{Method} & \textbf{RCV\textsubscript{D} (\%)} & \textbf{RCV\textsubscript{f} (\%)} & \textbf{RCV\textsubscript{D*} (\%)} \\
\midrule
Bayesian & 4.6 $\pm$ 0.7 & 55.8 $\pm$ 8.7 & 30.0 $\pm$ 6.4 \\
MDN      & \textbf{4.5} $\pm$ 1.0 & 53.5 $\pm$ 11.0 & \textbf{16.0} $\pm$ 4.1 \\
Gaussian & 4.7 $\pm$ 0.9 & 42.5 $\pm$ 9.4 & 21.6 $\pm$ 4.2 \\
MLP      & \textbf{4.5} $\pm$ 0.9 & \textbf{37.9} $\pm$ 8.6 & 20.0 $\pm$ 4.2 \\
\bottomrule
\end{tabular}
}
\caption{Median (MAD) of RCV values for $D$, $f$, and $D^*$ estimated in a ROI within the brain test set, using Bayesian inference, MDN, Gaussian model, and MLP.}
\label{tab:rcv}
\end{table}

\subsubsection{Analysis of the predicted uncertainties}

Since no ground truth is available in the in vivo test set, calibration and sharpness metrics cannot be computed; we therefore focus solely on image‐based uncertainty.  As for the phantom data, we calculated the median uncertainty values, disentangled into AU and EU components, for \(D\), \(f\), and \(D^*\).  The last row of Table~\ref{tab:combined_uq} summarizes these values for the Gaussian and MDN models.

In this dataset, Gaussian and MDN models show similar AU for \(D\) and \(f\), but the MDN yields lower AU for \(D^*\), consistent with the slight overconfidence observed in simulations.  Moreover, MDN produced lower EU across all parameters—especially \(f\) and \(D^*\)—indicating reduced heterogeneity in the posterior space between components of the ensemble.

\section{Discussion}

This study introduced a comprehensive framework for voxel-wise supervised IVIM parameter estimation, leveraging probabilistic regression and DE. Our primary contribution lies in extending traditional supervised learning approaches to not only provide accurate point estimates of IVIM parameters but also to quantify both AU and EU. Moreover, we assessed the reliability, in terms of calibration and sharpness, of these uncertainty estimates.

While previous works on UQ in other dMRI applications (e.g., \cite{jallais2024introducing,consagra2025deep,manzano2025uncertainty,karimi2023deep}) have employed probabilistic regression models—such as MDNs and normalizing flows—to estimate the posterior distribution of parameters, they have typically focused on quantifying only the AU counterpart within the distributional regression framework. In contrast, we extended the probabilistic regression setup through a DE-based approach, allowing us to marginalize the posterior over the model parameters. This key step enabled the simultaneous estimation of EU, providing a more comprehensive characterization of the overall predicted uncertainty.

Furthermore, we employed evaluation metrics widely used in the probabilistic regression literature \cite{kuleshov2018accurate,sluijterman2024evaluate} to assess the reliability of the predicted uncertainties, specifically evaluating the calibration and sharpness of the resulting predictive distributions. These metrics have been largely overlooked in medical image analysis probabilistic regression studies \cite{lopez2025uncertainty} and were not employed in aforementioned dMRI works. By bridging this methodological gap, our framework offers a robust and interpretable approach to UQ for IVIM parameter estimation. Importantly, the framework is generalizable and can be extended to other physical models by substituting the underlying data-generating model and adapting the network architecture accordingly.

\subsection{Simulated data}

Regarding the IVIM parameter estimation performance on simulated data, our probabilistic networks (Gaussian and MDN) and the non‑probabilistic MLP exhibited similar accuracy, yet all three significantly outperformed a state-of-the-art Bayesian IVIM fitting algorithm. These results are in line with other studies which used simulated phantoms as test set \cite{mastropietro2022supervised,casali2025physics}. Notably, differences between the two probabilistic models emerged in the RCV for $D^*$. At SNR=25, the MDN showed greater robustness (RCV=28.5\%) compared to the Gaussian model (RCV=37.2\%), and this trend persisted at SNR=50 (25.8\% vs 28.8\%). These findings suggest that the MDN’s capacity to model multi-modal or skewed posterior distributions yields more stable parameter estimates, in particular for those parameters that are more challenging to calculate in noisy conditions.

 The strategy we adopted to derive a point-estimate from the predicted probabilistic distribution was MAP estimation. Specifically, for each input voxel, we used the mean of the output Gaussian in the single Gaussian model, and for the MDN, we selected the mean of the component with the highest mixture weight (i.e., the mode of the mixture distribution). This effectively assumes that the most probable component provides the best representation of the prediction. However, alternative techniques-such as using the median of the predicted distribution-are also viable. The exploration of such alternatives is left for future work.

The values of uncertainty were normalized by the prior range of the synthetic parameters \cite{jallais2024introducing}, which allowed comparison across parameters with different physical units. These uncertainties were useful to provide an estimate of the possible confidence interval around the predicted values. Moreover, the use of MDNs, which allowed capturing multimodal distributions unlike using a single Gaussian parametrization, can enable the exploration of the full posterior distribution and supports further voxel-wise analysis by highlighting degeneracies, as done in Jallais et al.\cite{jallais2024introducing}.

Regarding the quantitative analysis of these uncertainties, our comprehensive calibration analysis revealed that the MDN generally provided better calibration for \(D\) and \(f\), meaning that its predicted uncertainty ranges accurately reflect the observed variability. These parameters estimates are better calibrated and sharper, as demonstrated by lower values of the PINAW$_{90\%}$ and by overall values of CRPS, which is a proper score that jointly evaluates calibration and sharpness. This aligns with the general rule of maximizing the sharpness of the predicted distirbution subjected to calibration, as expressed by Gneiting et al.\cite{gneiting2007probabilistic}. 
Regarding \(D^*\), the Gaussian model demostrated better calibration than MDN, where the latter showed a slightly overconfidence behaviour, which may be linked to its major flexibility to capture more complex, multimodal distributions. In such scenarios, the MDN's flexibility might attempt to model spurious noise patterns, leading to overfitting to noise and overly sharp, slightly miscalibrated uncertainty estimates. Regularization techniques might help in counteracting this effect on this parameter and will be explored in future works.

Regarding the disentanglement of the two uncertainty components, results revealed that the majority of uncertainty on simulations was AU, due to noise in the test set. AU decreased as expected with increasing SNR, whereas EU remained minoritary due to the fact that the simulation test set was similar to the training set, so these data are in-distribution for the NNs. The uncertainty was highest in \(D^*\), due to the fact that this parameter is particularly subject to noise because the contribution of the perfusion part to the overall signal decay is scarce and only relevant at low \(b\)-values. 

\subsection{In vivo data}

Since ground truth IVIM parameters are not available in vivo, the evaluation focused on the median values and the RCV within manually delineated "homogeneous" ROIs. Among the probabilistic approaches, the MDN achieved the lowest RCV for the \( D^* \) parameter in the in vivo dataset (16.0\%). The greater homogeneity in the MDN's estimation of \( D^* \) is also evident in Figure~\ref{fig:vivo}, where the Gaussian model exhibits regions with \( D^* \) values exceeding 0.15~mm\(^2\)/s, which are patterns not observed in the Bayesian or MDN estimations. For \( f \), the Gaussian model achieved a lower RCV (42.5\%) compared to the MDN (53.5\%).

The analysis of uncertainty revealed substantially higher values of EU and AU for the \( f \) and \( D^* \) parameters in vivo, compared to simulated data. This is consistent with the greater variability and noise characteristics inherent to real measurements. Although training simulations were designed to uniformly span the physiologically plausible range of IVIM parameters, the higher EU observed in vivo suggests that simulations may not fully capture the complexity of real acquisition conditions. For example, the presence of Alzheimer’s-related pathology, in our preclinical model, may have introduced additional complexity, including regions of impaired perfusion and microvascular damage. These pathological changes likely increased the heterogeneity of the real data in ways that were not fully anticipated by the training simulations. Other factors such as scanner-specific artifacts, bias field inhomogeneities, motion artifacts, and deviations from the assumed IVIM model can all introduce systematic discrepancies between synthetic and real signals.

While EU was generally lower on average compared to AU, as shown in Table~\ref{tab:combined_uq}, neglecting this component can lead to an underestimation of the total predictive uncertainty for specific regions. This can be seen in Figure~\ref{fig:vivo}. In the \( f \) maps, elevated \( f \) values are observed in the ventricular regions, which are filled with cerebrospinal fluid and are not expected to exhibit perfusion. In such regions, the IVIM signal primarily reflects diffusion, and due to low \( D^* \), the contributions of diffusion and perfusion may become indistinguishable \cite{zhang2019implicit}. These erroneous estimations are flagged by high AU values and, to a lesser extent, elevated local EU, which reinforces the importance of accounting for both uncertainty types when interpreting model outputs in complex anatomical regions.

\subsection{Possible future directions}
Future work could explore strategies for inspecting EU to flag out-of-distribution signals, potentially through training simulations that differ in range from those used during testing. Also, investigating the impact of simulated artifacts and bias field inhomogeneities—both plausible sources of EU—also represents a promising direction. Another possible approach may involve using separate neural networks for each IVIM parameter and to optimize calibration and sharpness on a per-parameter basis. Additionally, comparing the performance of MDN with normalizing flows \cite{papamakarios2021normalizing} within a DE framework may offer insights into their effectiveness for IVIM parameter estimation. Finally, exploring additional Bayesian Deep Learning approaches presents a promising direction for future research.

\section{Conclusions}

In conclusion, we present a comprehensive framework for the estimation of IVIM parameters using MDN combined with DE. Additionally, we introduce quantitative metrics to rigorously assess the reliability of uncertainty estimates, in terms of calibration and sharpness, while also disentangling both AU and EU contributions. These aspects are often overlooked in existing literature in this specific field. The integration of NNs with robust and complete UQ holds promise for facilitating the clinical adoption of the IVIM model, which traditionally suffers from noisy parameter estimation, particularly in the perfusion compartment. This approach can be especially valuable for radiologists, enabling them to interpret and flag unreliable estimates, representing a fundamental aspect in medical scenarios and in the presence of pathological conditions.

\section*{Acknowledgments}

This work was partially funded by the Italian MUR under the program PRIN 2022 (project NINFEA 2022MXW52Y - CUP B53C24006690006) 
and Fondazione Regionale per la Ricerca Biomedica (Regione Lombardia), project ID 1739635.

\subsection*{Author contributions}

\subsection*{Financial disclosure}

None reported.

\subsection*{Conflict of interest}

The authors declare no potential conflict of interests.

\subsection*{Code availability}

The code is publicly available at:
https://github.com/Bio-SimPro-Lab/comprehensive-framework-ivim.git

%\bibliography{MRM-AMA}%

\vfill\pagebreak

\end{document}